\documentclass[prl,twocolumn,floatfix,showpacs]{revtex4}
\usepackage{graphicx}
\usepackage{nccmath}
\usepackage{mciteplus}

\newcommand{\bpsi}{\boldsymbol{\psi}}
\newcommand{\bgamma}{\boldsymbol{\gamma}}

\begin{document}
\title{Nonlocal conductance reveals helical superconductors}
\author{B. B\'eri}
\affiliation{Theory of Condensed Matter Group, Cavendish Laboratory, J.~J.~Thomson Ave., Cambridge CB3~0HE, UK}
\date{February 2011}
\begin{abstract}
Helical superconductors form a two dimensional, time-reversal invariant topological phase characterized by a Kramers pair of Majorana edge modes (helical Majorana modes). Existing detection schemes to identify this phase rely either on spin transport properties, which are quite difficult to measure, 
or on  local charge transport, which  allows only a partial identification. 
Here we show that the presence of helical Majorana modes can be unambiguously revealed by measuring the nonlocal charge conductance. Focusing on a superconducting ring, we suggest two experiments that provide unique and robust signatures to detect the helical superconductor phase. 
\end{abstract}
\maketitle
 \
 \vspace{-0.65cm}

Topological phases of matter are at the forefront of current condensed matter research. The excitations in such a phase are gapped in the  bulk, but the nontrivial topology is indicated by the presence of robust gapless boundary modes.
There are various types of topological phases, which can be classified by the nature of the boundary modes they support (see Ref.~\onlinecite{Schnyder08,*HasanKaneRMP,*QiZhangRMP} for an overview). 
The most well known examples are provided by  integer quantum Hall (IQH) systems\cite{Klitzing,*TKNN} and their superconducting analogues, chiral superconductors\cite{Volovik88,*Volovik97,*Goryo1999}. These two dimensional systems realize time-reversal symmetry breaking  topological phases characterized by chiral edge modes. They are distinguished by the nature of these modes, which are ordinary fermions in IQH systems, and Majorana fermions in chiral superconductors\cite{Halperinedge,*Senthil00,ReadGreen,*Fendley07}.

The recent excitement about topological phases was triggered by the discovery of the quantum spin Hall (QSH) effect, because it realizes  a {\it time-reversal invariant} topological phase\cite{KaneMele05,*BZ06,*BHZ06}. QSH systems can be viewed as analogues of IQH systems,  with a Kramers pair of counterpropagating  fermion modes (helical modes) on their edge. The existence of the QSH phase was confirmed in experiments\cite{Konig07}. As IQH systems, chiral superconductors  also have time-reversal invariant counterparts, called helical superconductors\cite{Volhelical,*XHRZ09,Tanaka09,SatoFujimoto}. In this phase, the edge modes are helical Majorana fermions. Recent research shows that noncentrosymmetric superconductors provide a candidate for a solid-state realization\cite{Tanaka09,SatoFujimoto}. 
In order to experimentally demonstrate the existence of the helical superconductor phase, it is  important to develop detection schemes for observing  helical Majorana modes.

In the literature, there are two main strands of detection schemes. 
One direction focuses on spin transport properties\cite{Vor08,*Lu09,Tanaka09,SatoFujimoto,Mizuno2010,*Eschrig10}, but the experimental detection of spin transport is quite difficult in general. The other direction studies the effects of the helical edge modes on local charge transport, most notably on the  Andreev tunneling conductance\cite{Yok05,*Inio07,*Tanaka10,*Asano10,Mizuno2010,*Eschrig10}. 
While charge transport is much easier to measure, the Andreev tunneling spectrum only indicates that there are edge states below the superconducting gap, it does not expose their helical Majorana character. 

\begin{figure}[t]
\includegraphics[height=4cm]{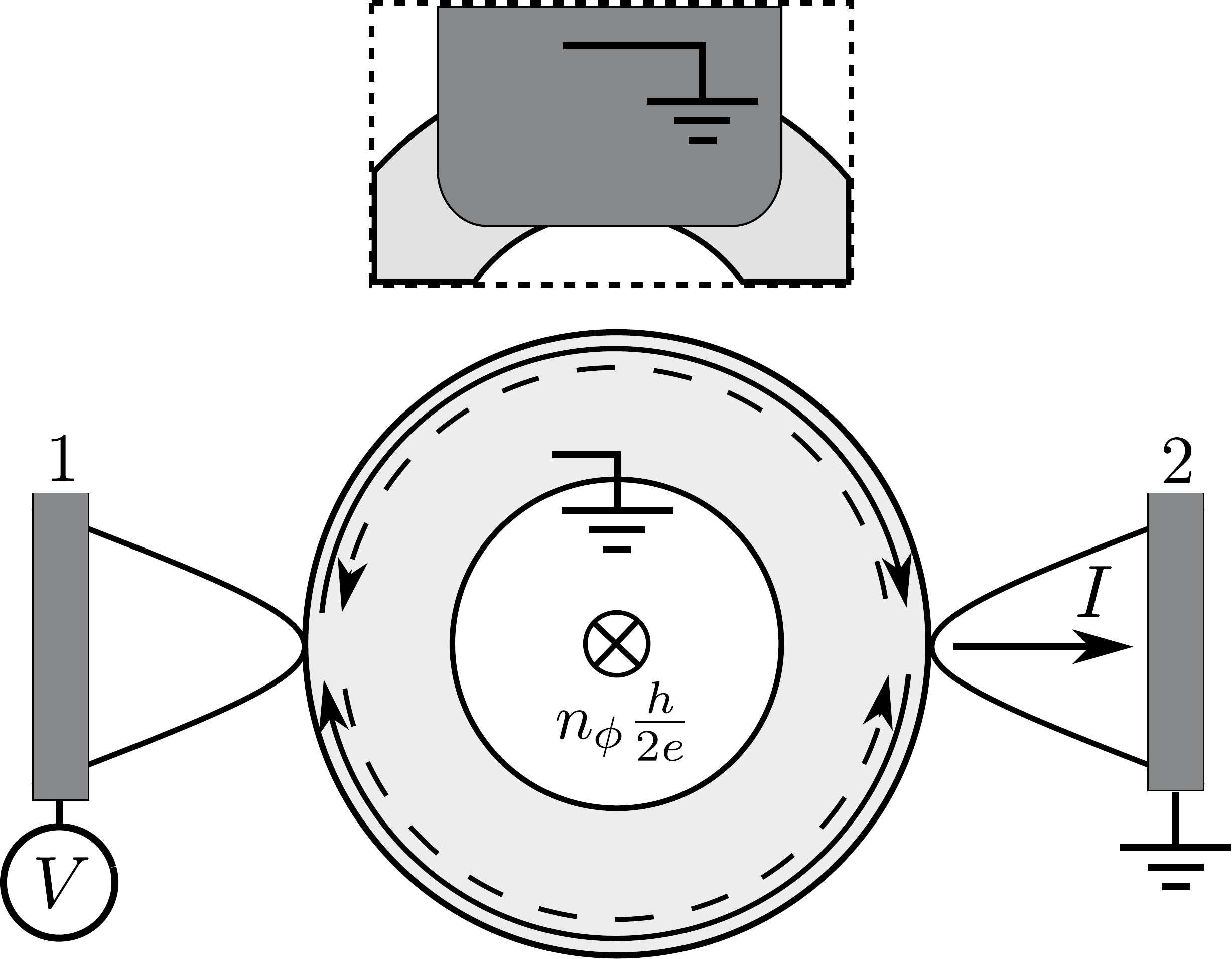}
\caption{Superconductor ring for detecting the helical superconductor phase. The normal-metal contact 1 at voltage $V$ injects charge into the  Majorana edge modes $\hat{a}$ (solid arrow) and $\hat{b}$ (dashed arrow) which is detected as an electrical current $I$ in contact 2.  
The conductance vanishes if a third, macroscopic contact (shown in the inset) is placed on either of the edges. An odd $n_\phi$  flux quanta leads to a Kramers pair of zero modes on the edge, changing the sign of the conductance.}
\label{fig:setup}
\end{figure}

In this paper we show that the charge conductance can be used to detect the existence and helical Majorana nature of the edge modes, if it is measured \emph{nonlocally}. 
To explain our considerations qualitatively, we first note that for temperatures and voltages well below the gap, any transport between two spatially separated contacts must take place via the edge modes. A fundamental property of Majorana edge modes at low energies is that one of them cannot transport charge, but a superposition of two can\cite{Akh09,*Fu09,Ser10}. These facts can be exploited using a setup in Fig~\ref{fig:setup}, where the (outer) edge of a helical superconductor ring is split into an upper and a lower segment by two normal-metal contacts. 
On the upper edge the Majorana mode $\hat{a}$ propagates from contact 1 to 2, while on the lower edge $\hat{b}$ does the same. 
Given that the contacts couple to a superposition of $\hat{a}$ and $\hat{b}$, the 
conductance $G$ between the contacts (i.e. the nonlocal conductance) is generally nonvanishing. While measuring a nonzero $G$ could mean several things (e.g. the presence of ordinary fermion edge modes or the lack of a bulk gap),  for a helical superconductor $G$ displays two unique features which allow the unambiguous detection of this phase.

 The first feature is the suppression of $G$ upon placing a third (grounded)
 macroscopic contact (illustrated in the inset of Fig.~\ref{fig:setup}) on either of the edges connecting 1 and 2. 
The role of this contact is to act as a perfect sink, thereby eliminating propagation along one edge from contact 1 to 2. Since now only one Majorana mode propagates from contact 1 to 2, $G$ vanishes. This effect directly demonstrates the helical Majorana nature of the edge modes: that the contact on either of the edges does the same shows that both edges are needed for charge transport.

A second possibility offered by our setup is to perform Majorana interferometry\cite{Akh09,*Fu09}, i.e.,  to switch the magnitude and sign of $G$ by controlling the  number of flux quanta in the helical superconductor ring of Fig.~\ref{fig:setup}. ($G$ can be negative because the outcoming quasiparticles in contact 2 can be holes, not only electrons, corresponding to Cooper-pairs  left behind in the superconductor.) One might worry at this point that the flux breaks time-reversal invariance, so it might spoil an essential element of the edge states. 
One of our key observations is that this is not so: flux quantization ensures that the flux enclosed by the edges is a multiple of $h/2e$, which is compatible with time reversal invariance of the edge physics\cite{Bar10,remaindernote}. The switching behavior is a characteristic signature of the appearance a zero mode in the scattering region. In the present case, an odd number of flux quanta leads to a Kramers pair of zero modes on the edge. That the zero modes switch the magnitude of $G$ is simply due to resonant tunneling. That they also switch the sign of $G$ follows from the  $\pi$ phase shift they induce between the upper and lower edges\cite{Akh09,*Fu09}. If without a zero mode, the Majorana fermions $\hat{a}=\hat{a}^\dagger$, $\hat{b}=\hat{b}^\dagger$ combine into an electron  according to $\hat{a}+i\hat{b}=\hat{\psi}^\dagger$ at the second contact, a zero mode changes this to a hole, $\hat{a}-i\hat{b}=\hat{\psi}$, which switches the sign of $G$.  Such an interferometric signal shows that there are Majorana modes going from 1 to 2  on both edges, indicating the helical Majorana nature of the edge states independently from the first feature.

We now turn to the quantitative analysis of the effects. 
As in Fig~\ref{fig:setup}, we take the superconductor and contact 2 to be grounded, and define the nonlocal conductance as $G=\partial I/\partial V$, with $I$ the current in contact 2, and $V$ the voltage in the first contact. 
If contact $\alpha$ ($\alpha=1,2$) supports $N_\alpha$ electron and hole modes, it is characterized by a scattering matrix
\begin{equation}
\begin{pmatrix}\bgamma\\\bpsi\end{pmatrix}_{\alpha,\text{out}}
=
S_\alpha
\begin{pmatrix}\bgamma\\\bpsi\end{pmatrix}_{\alpha, \text{in}},\quad
S_\alpha=\begin{pmatrix}
r_{\alpha} & t_\alpha\\
t^\prime_\alpha & r^\prime_\alpha
\end{pmatrix}
\end{equation}
relating the amplitudes $\bpsi=(e_1,\ldots,e_{N_\alpha},h_1\ldots h_{N_\alpha})^T$ of incoming electrons and holes on the normal side of the contact and the amplitudes $\bgamma_\alpha=(a_\alpha,b_\alpha)^T$ of incoming Majorana edge modes to the outgoing amplitudes. (Here and in what follows, we use the terms incoming and outgoing with respect to the discussed scattering region.) We work with temperatures and voltages much smaller than the superconducting gap and the inverse dwell time of the contacts, which allows to neglect the energy dependence of $S_\alpha$. Due to electron-hole and time-reversal symmetries, $S_\alpha$ satisfies\cite{AZ,Ser10}
\begin{equation}
\begin{pmatrix}
\openone_2 & \\
  & \Sigma_1
\end{pmatrix}
S_\alpha^*
\begin{pmatrix}
\openone_2 & \\
  & \Sigma_1
\end{pmatrix}
=S_\alpha,\quad s_2 S_\alpha^T s_2=S_\alpha
,\end{equation}
where $\Sigma_1=\sigma_1\otimes \openone_{N_\alpha}$ is the first Pauli matrix in electron-hole space, and $s_2$ the second Pauli matrix in spin-space. This amounts to the restrictions
\begin{equation}
r_{\alpha}=\chi_\alpha \sqrt{1-T_\alpha}\openone_2,\quad \chi_\alpha=\pm 1,
\end{equation}
and 
\begin{equation}
t_\alpha=\sqrt{T_\alpha}W_\alpha,\ W_\alpha W_\alpha^\dagger=\openone_2,\ W_\alpha^*\Sigma_1=W_\alpha.
\label{eq:tconstr}
\end{equation}
The transmission probability $T_\alpha$ measures the coupling strength between the contact $\alpha$ and the edge modes. Requiring that contact does not influence the edge modes in the uncoupled limit $T_\alpha\rightarrow 0$ sets $\chi_\alpha=1$.
At energy $E$, the upper and lower edges have scattering matrices
\begin{equation}
\bgamma_{\beta,\text{out}}
=
t_\beta(E)\openone_2
\bgamma_{\beta,\text{in}}
,\quad \beta=u,l,
\label{eq:tedge}\end{equation}
where the $\openone_2$ is again due to time-reversal invariance. Unitarity implies $|t_\beta(E)|\!=\!1$, and electron-hole symmetry requires \mbox{$t(-E)^*\!=\!t(E)$.}
The total transmission matrix relates incoming electron and hole amplitudes at contact 1 to outgoing ones at contact 2, $\bpsi_2\!=\!t_\text{tot}\bpsi_1$. It is given by
\begin{equation}
t_\text{tot}=\frac{\sqrt{T_1T_2}W_2^\prime t_M(E) W_1}{1-\sqrt{R_1R_2}t_u(E)t_l(E)},
\label{eq:ttotgen}
\end{equation}
where $t_M\!=\!\text{diag}(t_u,t_l)$, $W_2^\prime=s_2 W_2^T s_2$ and $R_\alpha\!=\!1-T_\alpha$. 
At temperature $T$, and in $e^2/h$ units, $G$ is given by\cite{Lam93}
\begin{equation}
G(eV)=\int_{-\infty}^\infty \left(-\frac{\partial f(E-eV)}{\partial eV}\right) g(E)dE
\label{eq:Ggen}
\end{equation}
with the Fermi function $f(\varepsilon)=[\exp(\varepsilon/k_BT)+1]^{-1}$, and 
$g(E)=\text{Tr}[t_\text{tot}^{ee}(t_\text{tot}^{ee})^\dagger]-\text{Tr}[t_\text{tot}^{he}(t_\text{tot}^{he})^\dagger]$. 
From $t_\text{tot}$ we obtain
\begin{equation}
g(E)=\frac{-T_1T_2g_1 g_2 \cos[\rho(E)]}{1-2\sqrt{R_1R_2}\cos[\theta(E)]+R_1R_2}.
\label{eq:gEgen}
\end{equation} 
Here we introduced $\theta(E)\!=\!\arg(t_lt_u)$ and \mbox{$\rho(E)\!=\!\arg(t_lt_u^*)$,} the sum and the difference of the phase shifts along the upper and lower edges, respectively. The factors $T_\alpha g_\alpha$ are contact conductances where
\begin{equation}
g_\alpha=2\sum_{j=1}^{N_\alpha} |v^{(\alpha)}_j|^2-1.
\label{eq:galpha}
\end{equation}
Here $v^{(\alpha)}_j$ is the first half row of $\frac{1}{\sqrt{2}}\left(\begin{smallmatrix}1&i\\1&-i\end{smallmatrix}\right)W_\alpha$. Physically $T_\alpha g_\alpha$ corresponds to the conductance between incoming electron and hole modes and the outgoing charged edge state combinations $\hat{a}\pm i\hat{b}$ at contact $\alpha$.
 Expressions (\ref{eq:ttotgen}-\ref{eq:galpha}) are the central formulas which will be used to quantitatively analyze the predicted effects. 

We first study the effect of a large contact (inset of Fig.~\ref{fig:setup}), taken to be on the upper edge for definiteness. In the presence of the contact, scattering on the upper edge becomes a three terminal problem. That the macroscopic contact acts as a perfect sink corresponds to decoupling of $\bgamma_{u,\text{in}}$ from $\bgamma_{u,\text{out}}$. In Eq.~\eqref{eq:ttotgen}, this replaces $t_u\rightarrow 0$, which, in effect, erases the second row of $W_\alpha$. In obtaining the conductance, this results in the replacement $g_\alpha\rightarrow 0$, which directly leads to $G=0$. The presence of the large contact thus suppresses the conductance as claimed in the qualitative discussion.  

To analyze the effect of flux quanta in the ring in Fig.~\ref{fig:setup}, we first establish that an odd number of them creates zero modes on the edges. We use the fact that for any helical superconductor, there exists a deformation which does not close the gap, respects time-reversal invariance, and decouples the system into two time-reversed copies of chiral superconductors\cite{Schnyder08}. Each of these chiral superconductors supports a chiral Majorana edge mode. (The two modes combine into a helical mode.) In the presence of an odd number of flux quanta, the edge of each copy supports a Majorana zero mode\cite{ReadGreen,*Fendley07}. Due to the   flux, the two zero modes are not time-reversed partners. However, flux quantization implies that the vector potential and the superconducting phase wind in such a way, that the edge theory is invariant under an antiunitary symmetry $\cal T$ combining time-reversal with a (large) gauge transformation. It is the presence of this $\cal T$-symmetry what is meant by saying that the quantized flux is compatible with the time-reversal invariance of the edge theory. Since $\cal T$ squares to $-1$, the eigenvalues of the edge Hamiltonian come in Kramers degenerate pairs. In addition, the edge theory has an electron-hole symmetry, dictating that the eigenvalues come in opposite pairs\cite{AZ}. Hence, the pair of zero modes cannot be removed\cite{Ivanov01}, even if the decoupled chiral superconductors are deformed back to the original helical superconductor. 

The presence of zero modes can be related to the edge scattering matrix Eq.~\eqref{eq:tedge}. The energy levels $E_n$ of the closed ($T_\alpha\rightarrow 0$) edge are determined by the secular equation
$t_u(E_n)t_l(E_n)=1$,
expressing the periodic boundary conditions. At zero energy, $t_{u,l}(0)=\pm 1$ due to electron-hole symmetry. For $n_\phi$  flux quanta we thus have 
\begin{equation}
t_u(0)t_l(0)=(-1)^{n_\phi+1}. 
\label{eq:tphi}
\end{equation}
For the zero bias, zero temperature conductance this leads to 
\begin{equation}
G(eV\rightarrow 0)=\frac{(-1)^{n_\phi}T_1T_2g_1 g_2}{1-(-1)^{n_\phi+1}2\sqrt{R_1R_2}+R_1R_2}. 
\label{eq:Gswitch}
\end{equation}
Importantly, the contact parameters $T_\alpha$, $R_\alpha$, $g_\alpha$ are independent of the flux through the ring\cite{cnfnote}. Eq.~\eqref{eq:Gswitch} thus expresses (the $eV,k_BT\rightarrow 0$ limit of) the second signature announced earlier, that the magnitude and sign of the nonlocal conductance can be switched by~$n_\phi$.

\begin{figure}
\includegraphics[width=\columnwidth]{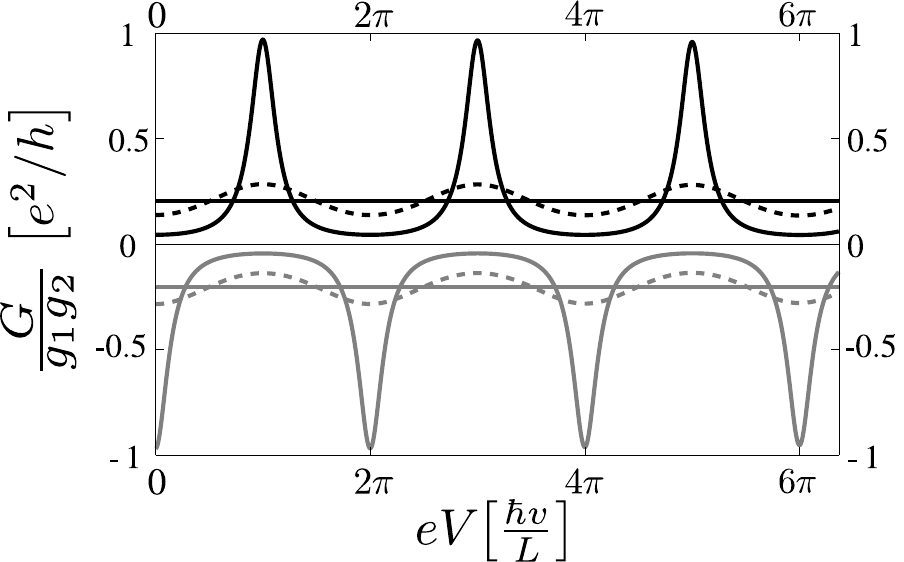}
\caption{$G(eV)$ for temperatures $T\!=\!0$ (solid lines with peaks), $k_BT\!=\!\hbar v/L$, (dashed lines), $k_BT\!=\!10\hbar v/L$ (horizontal lines). The parameters used are $T_1\!=\!0.3$, $T_2\!=\!0.4$, $\delta L/L\!=\!0.01$. Black (gray) curves correspond to $n_\phi$ even (odd). }
\label{fig:GV}
\end{figure}

To study the effect of finite temperatures and voltages, we consider a translation invariant edge with
\begin{equation}
t_l(E)=e^{i\phi_l(E)},\ \  t_u(E)=\xi e^{i\phi_u(E)},
\label{eq:tE}
\end{equation}
where $\phi_\beta(E)=EL_\beta/\hbar v$, with the edge velocity $v$ and the length $L_\beta$, and $\xi$ is the Berry phase for moving a Majorana mode around the edge. 
From Eq.~\eqref{eq:tphi} we have $\xi=(-1)^{n_\phi+1}$, which leads to 
\begin{equation*}
\theta(E)=EL/\hbar v+(n_\phi+1)\pi, \ \rho(E)=E\delta L/\hbar v + (n_\phi+1)\pi
\end{equation*}
in Eq.~\eqref{eq:gEgen}. Here $L\!=\!L_u+L_l$ is the total circumference and $\delta L\!=\!L_u-L_l$. 
At zero temperature, $G(eV)$ shows peaks with a period of the edge level spacing $2\pi \hbar v/L$, with slow oscillations of period $2\pi \hbar v/\delta L$ superimposed. In the tunneling limit $T_\alpha\ll1$, the peaks have width \mbox{$(T_1+T_2)\hbar v/L$} and height $4g_1g_2T_1T_2/(T_1+T_2)^2$. An odd $n_\phi$ shifts the peaks with $\pi \hbar v/L$ and switches their sign. This is the finite voltage form of the second signature, shown in Fig.~\ref{fig:GV}.

The effect of the temperature is to gradually smear out the peaks of $G(eV)$. Importantly, $G(eV)$ stays nonzero for  $\hbar v/L\!\!\ll\!\!k_BT\!\!\ll\!\!\hbar v/\delta L$, reaching a voltage independent value $G_\infty$ for high temperatures. (With careful positioning of the contacts, $\delta L\!\ll \!L$ can always be achieved.) For large coupling $T_\alpha\sim1$,  $G_\infty\sim(-1)^{n_\phi}g_1g_2$, while in the tunneling limit one has $G_\infty\!=\!(-1)^{n_\phi}g_1g_2T_1 T_2/(T_1+T_2)$. 
The behavior of $G(eV)$ in the zero and high temperature limits and for an intermediate temperature is shown in Fig.~\ref{fig:GV}.

The magnitude of the effects is determined by the contact parameters $T_\alpha$ and $g_\alpha$, with a large effect favoring $T_\alpha$, $g_\alpha$ of large magnitude. Optimizing the coupling strengths $T_\alpha\in[0,1]$ amounts to optimizing the transparency of the contacts.  The values of $g_\alpha\in[-1,1]$ depend on  the  Majorana mode superpositions coupled to incoming electrons. Coupling only to one of the Majoranas corresponds to $g_\alpha=0$.  
Since the Majorana degrees of freedom are intertwined with the spin degrees of freedom, using contacts with strong spin-orbit coupling enhances $|g_\alpha|$. 
Another route to optimize $g_\alpha$ is to place the superconducting ring on a QSH edge. This couples the Majorana modes to the QSH edge modes on the edge segment outside of the ring, which play the role of the electron and hole modes of the contacts. (The QSH edge modes are gapped under the superconductor, hence the QSH edge segment inside the hole of the ring is decoupled from the transport.) This setup corresponds to a $2\times 2$ $W_\alpha$, which is severely constrained by electron-hole symmetry\cite{Ber09a,*Ber09b,Akh09}, allowing only $g_\alpha=\pm1$. With suitable contacts and due to the robustness against temperatures $k_BT\!\gg\! \hbar v/L$, the effects should have sufficient magnitude and a reasonable temperature window for finding them in experiments.

Before concluding, it is worth mentioning an implication of the flux parity effect besides detecting  helical superconductivity. Consider, instead of a ring, a helical superconductor disk with a ferromagnetic layer at the center (i.e., a heterostructure replaces the hole in Fig~\ref{fig:setup}). The central heterostructure realizes a region with effective chiral $p$-wave pairing, and is one of the proposed architectures for topological quantum computation\cite{SatoFujimoto,Lee09}. Vortices trapping  $h/2e$ flux are proposed as the fundamental objects for storing quantum information. Detecting their number parity through the sign of $G$ can be a practical application, once a helical superconductor platform is available. 

To summarize, we have shown how to demonstrate the existence of a helical superconductor phase using nonlocal conductance measurements. The helical Majorana nature of the edge modes is revealed in two characteristic features, (i) the suppression of $G$ due to a large third contact, and (ii) switching the magnitude and sign  of $G$ with the number of flux quanta threading the helical superconductor ring. Both of these features survive thermal smearing for temperatures  $k_BT\!\gg\! \hbar v/L$, making them robust qualitative signatures to look for in the experimental search for helical superconductors. The second effect can also have future applications in helical superconductor based architectures for topological quantum computation.

I gratefully acknowledge discussions with  N.~R.~Cooper. 
This work was supported by EPSRC Grant EP/F032773/1.
\\

 \bibliographystyle{apsrevM}

\begin{mcitethebibliography}{40}
\bibitem[{rem()}]{remaindernote}
\bibinfo{note}{The edge states can be shielded from possible magnetic fields
  not trapped by the ring for example using additional superconducting elements
  (not shown in Fig.~\ref{fig:setup}).}\relax
\mciteBstWouldAddEndPunctfalse
\mciteSetBstMidEndSepPunct{\mcitedefaultmidpunct}
{}{\mcitedefaultseppunct}\relax
\EndOfBibitem
\bibitem[{cnf()}]{cnfnote}
\bibinfo{note}{One can always use a gauge, where the superconducting phase (mod
  $2\pi$) and the vector potential vanish at the contacts, independently of the
  flux.}\relax
\mciteBstWouldAddEndPunctfalse
\mciteSetBstMidEndSepPunct{\mcitedefaultmidpunct}
{}{\mcitedefaultseppunct}\relax
\EndOfBibitem
\end{mcitethebibliography}
\ifx\mcitethebibliography\mciteundefinedmacro
\PackageError{apsrevM.bst}{mciteplus.sty has not been loaded}
{This bibstyle requires the use of the mciteplus package.}\fi

\end{document}